\documentstyle[12 pt,fleqn]{article}
\newcommand{\no}{\nonumber\\}
\newcommand{\be}{\begin{equation}}
\newcommand{\ee}{\end{equation}}

\newcommand{\ba}{\begin{eqnarray}}
\newcommand{\ea}{\end{eqnarray}}
\newcommand{\ci}[1]{\cite{#1}}
\newcommand{\bi}[1]{\bibitem{#1}}
\newcommand{\la}[1]{\label{#1}}

\def\gl#1{(\ref{#1})}
\def\tr#1{\mbox{\rm tr}\left(#1\right)}

\newcommand{\AmS}{{\protect\the\textfont2
  A\kern-.1667em\lower.5ex\hbox{M}\kern-.125emS}}

\begin{document}
\begin{center}
{\large\bf The Extended Chiral Quark Model confronts QCD}
\footnote{Talk given at the Euroconference QCD99, Montpellier, July 1999.}
\footnote{This work is supported by EU Network EURODAPHNE,  CICYT grant
AEN98-0431 and CIRIT grant 1998SRG 00026.
A.A. is
supported by the Generalitat de Catalunya (Program PIV 1999) 
and partially by RFBR grant 98-02-18137.}\\

\vspace{1cm}

{\bf A. A. Andrianov, D. Espriu and R. Tarrach}\\
\vspace{0.5cm}
Departament d'Estructura i Constituents de la Mat\`eria,
 Universitat de Barcelona,\\
 Diagonal 647, 08028 Barcelona, Spain\\
\end{center}

\vspace{2cm}

\begin{abstract} 
We discuss the truncation of low energy effective
action of QCD below the chiral symmetry breaking (CSB) scale,  including
all operators of dimensionality less or equal to 6 which can be built with
quark and chiral fields. 
We perform its bosonization in the scalar,
pseudoscalar, vector and axial-vector channels in the large-$N_c$
and leading-log approximation.
Constraints
on the coefficients of the effective lagrangian are
derived from the requirement of Chiral Symmetry Restoration (CSR) at
energies above the CSB scale in the scalar-pseudoscalar 
and vector-axial-vector channels, from
matching to QCD at intermediate scales, and by fitting
some hadronic observables. In this truncation two types of
pseudoscalar states
(massless pions and massive $\Pi$-mesons), as well as a scalar, vector and
axial-vector
one arise as a consequence of dynamical chiral symmetry
breaking. Their masses and coupling constants
as well as a number of chiral structural
constants
are derived.
A reasonable fit of all parameters 
supports a relatively heavy scalar meson (quarkonium) with the
mass $\sim 1$ GeV and a small value of axial pion-quark coupling
constant $g_A \simeq 0.55$.
\end{abstract}

\vfill
\noindent
UB-ECM-PF 99/14\\
\noindent
September 1999\\
\noindent
\newpage
\section{Introduction}
The basic idea of the Extended Chiral Quark Model (ECQM) \ci{aet}
consists in using the degrees
of freedom which are relevant at each energy scale.
It is 
built in
terms of colored current
quark fields $\bar q_i(x), q_i(x)$ with momenta restricted to be below the
CSB scale $\Lambda_{CSB} \sim
1.3$ GeV,
and colorless chiral fields
$U(x) = \exp \left(i\pi(x)/ F_0\right)$  
which are  $SU(N_F)$ matrices (herein $N_F = 2$) and which
appear below $\Lambda_{CSB}$. The quarks are
endowed with a  `constituent' mass $M_0$ multiplied by
the chiral field $U(x)$ without
manifestly breaking chiral symmetry. 
The information
on modes with momenta larger than $\Lambda_{CSB}$ as well as
the effect of residual gluon interactions
is contained
in the coefficients of the effective lagrangian. The ECQM
truncation of QCD effective action
happens to be an extension of both the  chiral quark model
and the
Nambu-Jona-Lasinio one.

The  external
sources are included into the QCD quark lagrangian
in order to compute the correlators of corresponding quark currents
\be
 \hat D \equiv i  \gamma_\mu(\partial_\mu+ \bar V_\mu  + \gamma_5\bar A_\mu) +
 i (\bar S + i \gamma_5 \bar P), 
\ee
where $\langle S\rangle = m_q$,  matrix of current
quark masses.

The low-energy effective lagrangian
 ${\cal L}_{ECQM}$ is built to be invariant
under left and right $SU(2)$  rotations,
of quark, chiral and external fields.

It is convenient to introduce the `rotated', `dressed' or `constituent' quark
fields 
\be
Q_L \equiv \xi q_L,\qquad Q_R \equiv \xi^\dagger q_R,\qquad  \xi^2\equiv U,
\ee
which transform nonlinearly under\\  $ SU_L(2) \bigotimes SU_R(2) $
but identically for left and right quark components.

Changing to the `dressed' basis implies the following replacements
in the external vector, axial, scalar and pseudoscalar sources
\ba
\bar{V}_\mu  \to  v_\mu &=& \frac12 \left( \xi^\dagger \partial_\mu \xi -
\partial_\mu \xi  \xi^\dagger +  \xi^\dagger \bar V_\mu \xi +
\xi \bar V_\mu\xi^\dagger - \xi^\dagger \bar A_\mu \xi +
\xi \bar A_\mu \xi^\dagger\right),\no
\bar{A}_\mu  \to  a_\mu &=& \frac12
\left( - \xi^\dagger \partial_\mu \xi -
\partial_\mu \xi \xi^\dagger -  \xi^\dagger \bar V_\mu \xi +
\xi \bar V_\mu\xi^\dagger + \xi^\dagger \bar A_\mu \xi +
\xi \bar A_\mu \xi^\dagger\right),\no
\bar{\cal M}  \to  {\cal M} &=& \xi^\dagger \bar{\cal M} \xi^\dagger .
\ea
In these variables the relevant part of ECQM action can be represented as, 
\be
{\cal L}_{ECQM} = {\cal L}_{ch} + {\cal L}_{\cal M} + {\cal L}_{vec},
\ee
where ${\cal L}_{ch}$ accumulates the interaction of chiral fields and
quarks in the chiral limit in the presence of vector and  axial-vector
external fields, 
${\cal L}_{\cal M}$ extends the description 
 for external scalar and pseudoscalar fields and, in particular,
for massive quarks,
${\cal L}_{vec}$ contains 
operators generating
meson states in vector and axial-vector channels. 

In more detail\footnote{
In \gl{lchir} we have retained numerically the most important operators
and only those  four-quark vertices which
induce a scalar isosinglet and pseudoscalar
isotriplet meson states.}
\ba 
{\cal L}_{ch}&=& 
i\bar Q \left( \not\!\! D
 +  M_0 \right) Q
 -\frac{f_0^2}{4}{\rm tr}(a_\mu^2)\no
&&+\,\frac{G_{S0}}{4N_{c} \Lambda^2}\,
(\bar{Q}_L Q_R  +
\bar{Q}_R Q_L)^2  
- \frac{G_{P1}}{4N_{c} \Lambda^2}\,
( -  \bar{Q}_L \vec\tau Q_R
+  \bar{Q}_R  \vec\tau Q_L)^2, \la{lchir}
\ea
where
\be
 Q \equiv Q_L + Q_R,\qquad
\not\!\! D \equiv \not\!\partial  + \not\! v - \gamma_5 \tilde g_A 
\not\! a , 
\ee
with  the `bare' pion decay constant $f_0$ and the `bare' 
axial coupling $\tilde g_A 
\equiv 1 - \delta g_A$. These  `bare' contributions to the
chiral effective lagrangian are renormalized
after integration over low-energy  quark

The massive part ${\cal L}_{\cal M}$ looks as follows
\ba
 {\cal L}_{\cal M} &=&
i (\frac12 + \epsilon) \left(\bar Q_R {\cal M}
 Q_L + \bar Q_L  {\cal M}^\dagger  Q_R \right)\no 
&& + i (\frac12 - \epsilon) \left( \bar Q_R {\cal M}^\dagger  Q_L
+  \bar Q_L  {\cal M} Q_R\right)\no
&& +  \tr{  c_0\left({\cal M}  + {\cal M}^\dagger\right) 
 + c_5 ({\cal M} +{\cal M}^\dagger)a_\mu^2 
+ c_8 \left({\cal M}^2 + \left({\cal M}^\dagger\right)^2\right)}  ,\ea
where the chiral couplings $c_0,c_5,c_8$ 
 are  `bare', different from \ci{gl} and
their physical values  are controlled by the CSR rules (see below).

The chiral invariant quark self-interactions in the vector and  axial-vector
channels, ${\cal L}_{vec}$, are
\ba
{\cal L}_{vec} &=& - \frac{G_{V1}}{4N_c \Lambda^2} \bar Q \vec\tau \gamma_\mu
Q \bar Q \vec\tau \gamma_\mu Q 
 -
\frac{G_{A1}}{4N_c \Lambda^2} \bar Q \vec\tau \gamma_5\gamma_\mu Q
\bar Q \vec\tau \gamma_5\gamma_\mu Q \no
&&+ c_{10} \tr{U \bar L_{\mu\nu} U^\dagger \bar R_{\mu\nu}}. 
\ea
Their inclusion leads to the appearance of
 vector and axial-vector isotriplet meson resonances .
$c_{10}$ is a `bare' chiral coupling.

In total  the effective action suitable for derivation of two-point 
correlators contains 13 parameters to be determined by matching to QCD:
$M_0, \Lambda$(cutoff), the bare chiral constants $ f_0, c_0, c_5, c_8, c_{10}$,
the axial pion-quark coupling $\tilde g_A$, the mass asymmetry
$\epsilon$ and the four-fermion coupling constants 
$G_{S0}\not= G_{P1}, G_{V1}\not= G_{A1}$. 

\section{Bosonization} 

We incorporate auxiliary fields $\Phi$ in the scalar and pseudoscalar
channels, $\Sigma, 
 \Pi^a $, and in the vector and
axial-vector channel,
$i W^{(\pm)a}_\mu $, 
and replace the
four-fermion operators
\be
\frac{G_{C}}{4N_c \Lambda^2}\bar{Q}\Gamma Q \bar{Q}\Gamma Q;\,\,
\Gamma = 1; i\gamma_5 \tau^a; \gamma_\mu\tau^a; \gamma_5 \gamma_\mu\tau^a,
\la{bos1} 
\ee
by
\be
i \bar{Q}\Gamma \Phi  Q
+ N_{c} \Lambda^2\frac{\Phi^2}{G_C};\quad C = S0; P1; V1; A1, \la{bos2} 
\ee
with an integration over new variables (see  
\ci{aet,ae}). 

Due to vacuum polarization effects (quark loops) the auxiliary
fields obtain kinetic terms and propagate, i.e. interpolate resonance states. 
One fulfills the confinement requirement for
a finite number of resonances if one retains only that part
of the quark loop which contains the leading logarithm of the cutoff $\Lambda$.
In this approach one coherently 
neglects both the threshold part of quark loop (`continuum') and
(the infinite number of)  heavier resonance poles. This is supported by
 the large-$N_c$ approximation which associates all momentum dependence in the
bosonized effective action  solely with meson resonances.

The actual value of
constituent mass $<\Sigma> = \Sigma_0$  is described
by the mass-gap equation
\be
\frac{ \Lambda^2}{G_{S0}}\left(\Sigma_0 - M_0 \right)
= - \frac{\Sigma^3_0}{4\pi^2} \ln\frac{\Lambda^2}{\Sigma_0^2}
\equiv \Sigma^3_0 I_0. \la{msg2}
\ee 
Therefrom it is evident that the natural scale for the four-fermion
interaction is given rather by $\Sigma_0$ than by $\Lambda$ and it is
useful to redefine the related coupling constants:
$\bar G_C = G_{C}I_0 \frac{\Sigma_0^2}{\Lambda^2}$ for characterizing
the weak coupling regime by
$\bar G_C \ll 1$.

The leading-log part of the quark loop allows to find analytically both the
mass spectrum and decay coupling constants of pions, heavy pions,
scalar, vector and axial-vector resonances.\\ 
In particular,
the physical axial coupling in pion-quark vertex is
$ g_A = \tilde g_A / (1 + \bar G_{A})$.\\
The masses of vector mesons are evaluated to be
\be
m_V^2
= \frac{6\Sigma_0^2}{\bar G_V},\quad m_A^2 = 
6 \Sigma_0^2 \frac{1 + \bar G_A}{\bar G_A}. \la{mV}
\ee
Among others, their coupling constants to
external vector fields are of main importance
\be
f_V = \frac{N_c I_0}{6} ;\qquad f_A = g_A f_V . \la{fva}
\ee 
The pion decay constant  can be
found by taking into account the bare pion kinetic term \gl{lchir}
\be
F_0^2 =
f^2_0 + N_c \Sigma^2_0 I_0 g_A \tilde g_A. \la{fpi}
\ee
The pion mass is set by the quark condensate
\be
C_q= \left(2 c_0 + \frac{N_c}{4 \pi^2}
\Sigma^3_0 \ln\frac{\Lambda^2}{\Sigma_0^2}\right) \equiv - B_0 F_0^2,
\ee
according to the Gell-Mann-Oakes-Renner formula, $m^2_\pi = 2 m_q B_0$ and
the masses of the $u,d$ quarks are taken equal for simplicity.

Respectively the heavy $\Pi$ mass is found to be
\be
m^2_{\Pi} =
\frac{ 2 \Sigma_0^2 \widetilde g_A}{\delta^2 g_A} 
(\frac{1}{\bar G_P} + 1),\quad \delta \equiv \frac{F_0}{f_0}. \la{masPi} 
\ee
The weak decay coupling constant for heavy $\Pi$ meson reads
\be
F_{\Pi} = F_0 d_1
\frac{ m^2_\pi}{m^2_\Pi(0)};\quad
d_1 =  \frac{\sqrt{1 - \delta^2} }{\delta} \left(
\frac{2 \Sigma_0\epsilon}{\bar G_{P} g_A B_0}
+1\right). \la{fPi}
\ee
The scalar meson mass is obtained in the form
\be
m^2_\sigma = 2 \Sigma_0^2 (\frac{1}{\bar G_S} + 3). \la{sigmas}
\ee
The matching to QCD yields further relations.

\section{CSR matching}
Let us exploit the constraints based
on chiral symmetry restoration at QCD at high energies.
We focus on two-point correlators of colorless 
quark currents
\be
\Pi_C (p^2) = \int d^4x \exp(ipx)
\langle T\left(\bar q\Gamma q (x) \bar q \Gamma q
(0)\right)\rangle, 
\ee
with the notations \gl{bos1} and \gl{bos2}.
In the chiral limit the scalar correlator  and the pseudoscalar one 
  coincide at all orders
in perturbation theory and also at leading order in the non-perturbative
O.P.E.\ci{svz}
(see also \ci{aet,zap}).The same is true for the difference 
between the vector
and axial-vector  correlators \ci{ppr}.
As the above differences decrease rapidly with increasing momenta, one can
expect that the lowest lying resonances included into ECQM will 
successfully saturate  the constraints from
CSB restoration.

In 
the scalar channel one obtains the following sum rules
\ba
&& c_8+\frac{N_c\Sigma_0^2 I_0}{8\bar G_{S}}
-\frac{4\epsilon^2 N_c \Sigma_0^2 I_0}{8\bar G_{P}}=0,\la{srul1}\\
&& Z_\sigma = Z_\pi + Z_\Pi, \qquad  Z_\pi = 4 B_0^2 F_0^2,\la{srul2}\\
&& Z_\sigma  m^2_\sigma -  Z_\Pi  m^2_\Pi \simeq  
24 \pi\alpha_s C_q^2 \sim 0, \la{srul3}
\ea
where 
$ Z_\sigma, Z_\pi,  Z_\Pi$ stand for the residues
in resonance pole contributions in the scalar and pseudoscalar correlators.
The first relation fixes unambiguously the bare constant $c_8$,
the last one is essentially 
saturated by heavy pion parameters. As result of CSR sum rules one
determines the chiral constant \ci{gl}
\be
L_8 \simeq 
\frac{F_0^2}{16}\left(\frac{1}{m^2_\sigma} + \frac{1}{m^2_\Pi}\right),
\la{L-8}
\ee
as well as the asymmetry
\be
2\epsilon   =\frac{\bar G_P}{\bar G_S} \sqrt{\frac{g_A}{\tilde g_A}}
\left( - \beta\sqrt{1 - \delta^2} \pm  \delta \sqrt{1 -
\beta^2}\right),
\la{beta}
\ee
where $\beta \simeq \sqrt{1 - (m^2_\sigma/m^2_\Pi)}$ and $\delta = f_0/F_0$.

In the vector channel one derives the relations 
\ba
&&c_{10} = 0, \la{vsr1}\\
&&f_V^2 m_V^2 = f_A^2 m_A^2 + F_0^2,\la{vsr2}\\
&&f_V^2 m_V^4 = f_A^2 m_A^4,  \la{vsr3}
\ea
where the two last ones represent the Weinberg sum rules.
With the help of the first relation one obtains the chiral constant\ci{gl,ppr}:
\be
L_{10} = \frac14 \left(f_A^2 - f_V^2\right).
\ee
From the second one and eq.\gl{fva} we find
\be
f_V^2 = \frac{F_0^2}{m_V^2 (1 - g_A^2 \xi)}
= \frac{N_c I_0}{6},\qquad
 \xi = 
\frac{m^2_A}{m^2_V}, \la{vsum1} 
\ee
and from the last one and eq.\gl{fva} it follows that
\be
 \xi = 
\frac{m^2_A}{m^2_V} = \frac{1}{g_A}. \la{lsr}
\ee

The last QCD requirement we adopt concerns the CSR 
for the three-point correlator 
of one scalar and two axial-vector currents \ci{aet}. It determines
eventually $c_5$ and the chiral constant $L_5$ \ci{gl}
\be
c_5 \simeq 0; \qquad  L_5 \simeq 
\frac{N_c \Sigma_0 I_0 g^2_A}{8 B_0(1 + 3 \bar G_S)}.\la{L5}
\ee
 
\section{Fit and discussion}

Let us  specify the input parameters.
As such we take
$F_0 = 90\ {\rm MeV}$,  $m_\pi^2 = 140\ {\rm MeV}$. We adopt \ci{gl,dn}
$\hat m_q (1\ {\rm GeV}) \simeq 6 \ {\rm MeV}$,
$B_0 (1\ {\rm GeV})\simeq 1.5\ {\rm GeV}$,
and engage the phenomenological value for the heavy pion mass
$m_\Pi \simeq 1.3$ GeV \ci{pdg}. We also take the vector and axial-vector meson
masses,
$m_\rho = 770\ {\rm MeV}$ and  $m_{a1} \simeq  1.2$
MeV, as known parameters. Then the parameter $\xi \simeq 2.4$. 

Let us perform now an optimal fit  applying in the vector channel only 
\gl{vsr1} and \gl{vsr2}. For $m_\sigma
\simeq 1$ GeV one finds $\beta \simeq
0.64$ and $L_8 \simeq 0.8\times 10^{-3}$. 
For $g_A =0.55$ one obtains
$L_5 =  1.2 \times 10^{-3}$ ($L_5, L_8$ to be compared with \ci{gl}) 
and $\Sigma_0 \simeq 200$ MeV.
 Therefrom 
one derives that $\bar G_V \simeq 0.25,\, \bar G_A \simeq 0.2, \,
\tilde g_A \simeq 0.66$. With these values, $I_0 \simeq 0.1$ and
$\Lambda \simeq 1.3$ GeV. Then the bare pion coupling $f_0
\simeq 62$ MeV and for the rest of the parameters we find:
$\delta \simeq 0.7$, $\bar G_S \simeq 0.11$, $\bar G_P \simeq 0.13$,
 and either
$\epsilon  \simeq 0.05$ or $\epsilon  \simeq - 0.51$. We see that indeed
the four fermion coupling constants $\bar G_S$ and $\bar G_P$ as well as
 $\bar G_V$ and $\bar G_A$ are slightly different and their values $\ll 1$
signifying the weak coupling regime. We remark that for the value 
$g_A =0.55$ the last sum rule \gl{lsr} is imprecise: 2.4 vs. 1.8 .  

The vector
and axial vector couplings are $f_V = 0.22$ and $f_A = 0.12 $ to be
well compared with the
experimental values \ci{pdg} from  the electromagnetic decays of $\rho^0$
 and $a_1$ mesons. 
Two more predictions can be obtained: 
$F_\Pi = 0.8\times 10^{-2} F_\pi, \qquad
F_\sigma =\frac{\sqrt{Z_\sigma}}{2 B_0} = 1.6 F_0$.      
These constants
are not yet experimentally measured.

Thus we have estimated all parameters of the ECQM effective lagrangian
and made certain predictions. 
We conclude that the ECQM supplied with the CSR matching conditions 
proves to be a 
systematic way 
to describe hadron properties at low and intermediate
energies starting from QCD. 

An alternative scheme exists for modelling 
the QCD effective action at intermediate energies
which is based on manifestly chiral invariant,
quasilocal many-quark interaction\ci{ava}. Like the
simple NJL model
it exploits the hypotetical CSB 
mechanism due to strong attraction in
scalar channels and yields  a rather  light scalar meson.

\noindent
{\bf D.Becirevic} (Orsay): {\it Could you comment on why you did not use
the last sum rule in the vector channel? How its inclusion may affect
the scalar meson mass?}\\
{\bf A.A.Andrianov}: {\it We, in fact, have performed  the fit employing
 the sum rule \gl{lsr}. As a result, the mass of axial-vector meson
comes out to be too low, 1 GeV or less, 
other parameters are changed 
slightly: $g_A$ grows up and $\Sigma_0$ decreases. Thus we have 
disfavoured  \gl{lsr} not being
satisfied with such a large discrepancy between physical and
large-$N_c$ values for $a1$ mass.
As to the scalar 
meson its mass is governed by the scalar sum rules and the chiral
constant $L_8$ and thereby is not affected by addition or neglection of  
\gl{lsr}.}

\begin{thebibliography}{9}
\bibitem{aet}  A. A. Andrianov, D. Espriu and R. Tarrach, Nucl. Phys. 
B533 (1998) 429.
\bi{gl} J. Gasser and H. Leutwyler, Nucl. Phys. B250 (1985) 465.
\bi{ae}  A. A. Andrianov and D. Espriu, hep-ph/9906459.
\bi{svz} M. A. Shifman, A. I. Vainstein and V. I. Zakharov,
Nucl. Phys. B147 (1979) 385, 448.
\bi{zap} A. A. Andrianov and V. A. Andrianov, hep-ph/9705364.
\bi{ppr} S. Peris, M. Perrottet and E. de Rafael, JHEP, 9805 (1998) 011.
\bi{dn} H. G. Dosch and S. Narison, Phys. Lett. B417 (1998) 173.
\bi{pdg} Particle Data Group: C. Caso et al., European Phys. J. C3 (1998) 1.
\bi{ava} A. A. Andrianov and V. A. Andrianov,
Int. J. Mod. Phys. A8  (1993) 1981;\\
hep-ph/9309297;\quad
 Nucl. Phys. Proc. Suppl.
39BC (1995) 257.
\end{thebibliography}
\end{document}